\documentclass[prl, amsmath, amssymb, twocolumn, showpacs]{revtex4}

\usepackage{graphicx}
\usepackage{color}
\usepackage{hyperref} 
\usepackage[labelformat=simple]{subfig}

\hypersetup{
pdfauthor={Rolf W. Reinthaler, Patrik Recher and Ewelina M. Hankiewicz},
pdftitle={Proposal for an all-electrical detection of crossed Andreev reflection in topological insulators},
pdfsubject={Publication in theoretical condensed matter physics},
pdfproducer={}
}

\newcommand{\ket}[1]{\left\vert #1 \right\rangle}
\newcommand{\bra}[1]{\left\langle #1 \right\vert}

\newcommand{\imag}{\mathrm{i}}
\newcommand{\ek}{\varepsilon(k)}
\newcommand{\Mk}{\mathcal{M} (k)}
\newcommand{\UnitM}{\mathbb{I}}
\newcommand{\rd}{\mathrm{d}}
\newcommand{\unit}[1]{\;\mathrm{#1}}

\newcommand{\Ef}{E_{\text{F}}}
\renewcommand{\Im}{\mathrm{Im}}
\renewcommand{\Re}{\mathrm{Re}}
\renewcommand{\vec}[1]{\boldsymbol{#1}}

\begin{document}
\date{\today}

\title{Proposal for an all-electrical detection of crossed Andreev reflection in topological insulators}
\author{Rolf W. Reinthaler$^{1}$, Patrik Recher$^{2}$ and Ewelina M. Hankiewicz$^{1}$}
\affiliation{
$^{1}$ Institute for Theoretical Physics and Astrophysics, University of W\"urzburg, D-97074 W\"urzburg, Germany
 \\
		$^{2}$ Institute for Mathematical Physics, Technical University Braunschweig, D-38106 Braunschweig, Germany}

\pacs{	72.25.Dc, 
		73.23.Ad, 
		74.45.+c  
	 }

\begin{abstract}
Using a generalized wave matching method we solve the full scattering problem for quantum spin Hall insulator (QSHI) - superconductor (SC) - QSHI junctions. We find that for systems narrow enough so that the bulk states in the SC part couple both edges, the crossed Andreev reflection (CAR) is significant and the electron cotunneling (T) and CAR become spatially separated. We study the effectiveness of this separation as a function of the system geometry and the level of doping in the SC. Moreover, we show that the spatial separation of both effects allows for an all-electrical measurement of CAR and T separately in a 5-terminal setup or by using the spin selection of the quantum spin Hall effect in an H-bar structure.
\end{abstract}

\maketitle

\begin{figure}
	\subfloat[]{\includegraphics[width=0.7\linewidth]{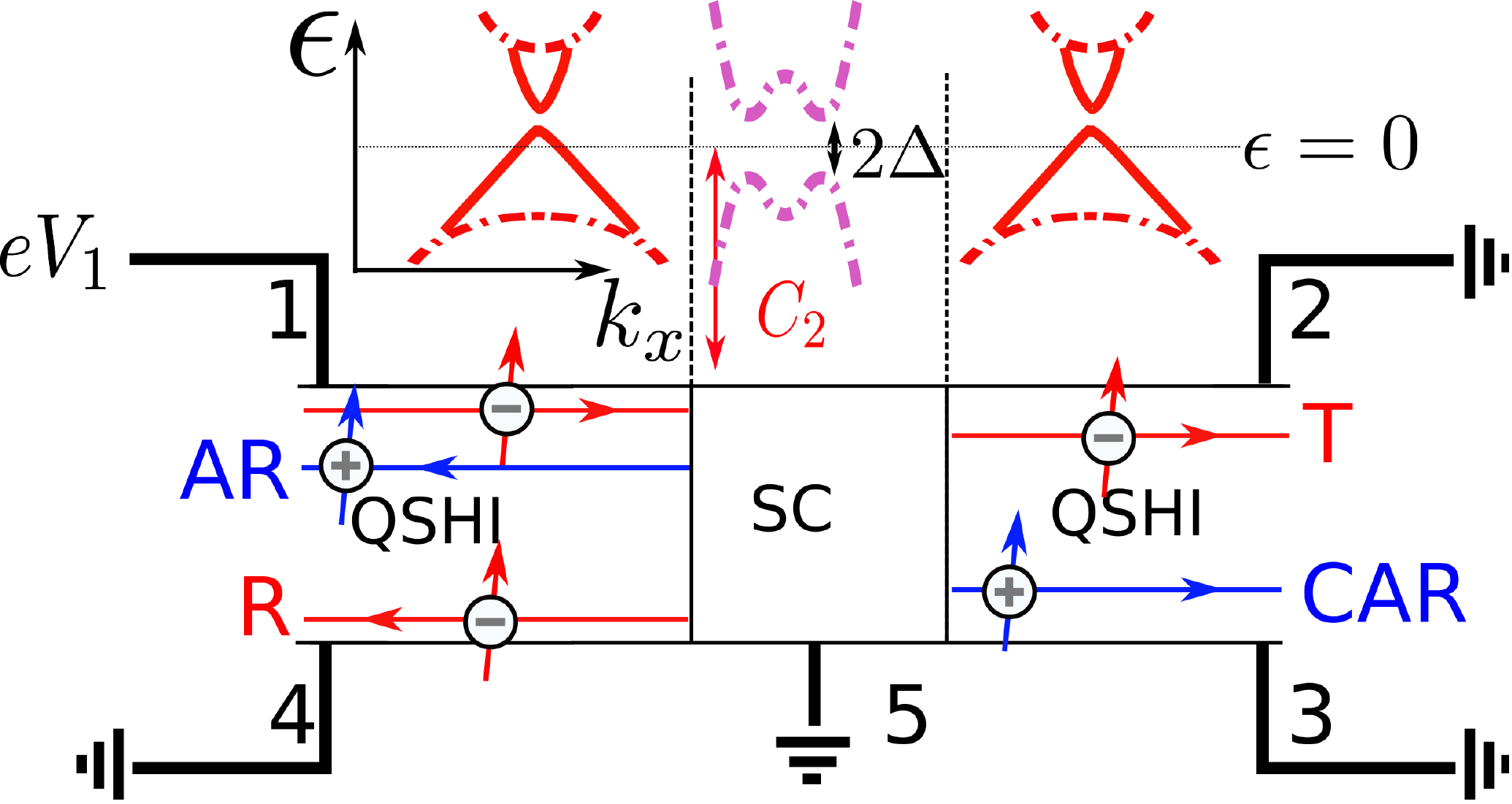}\label{fig:setupA}}\\
	\subfloat[]{\includegraphics[width=0.7\linewidth]{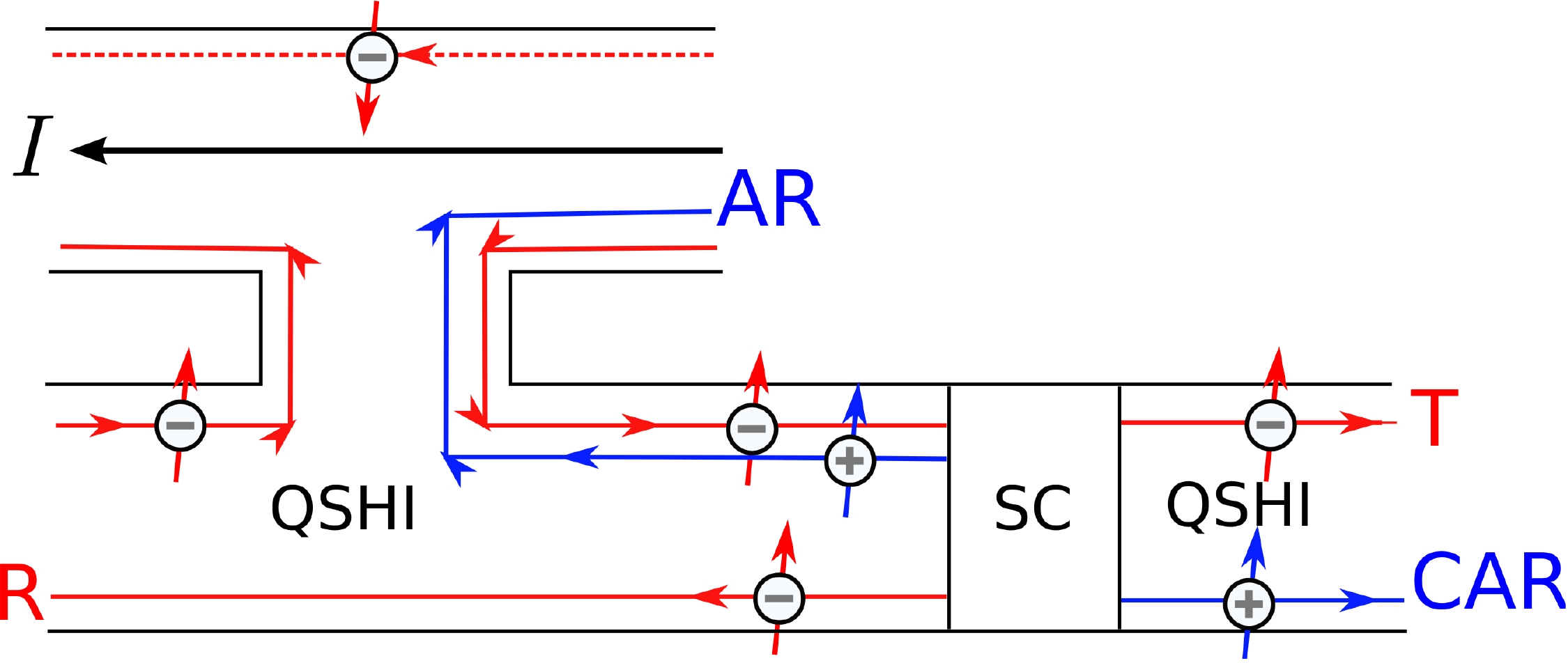}\label{fig:setupB}}
	\caption{(a) 5-terminal setup. We consider a QSHI-SC-QSHI junction. Electron (hole) edge states are indicated by red (blue) lines. Above each region we show the corresponding band structure schematically. Solid lines indicate edge states, dashed lines are bulk solutions. For simplicity only electron states are shown in the leads. Holes follow from inversion around $\epsilon = 0$. The SC is always grounded. (b) Spin selection via the H-bar structure of a QSHI. Driving a current $I$ in the upper leg of the structure leads to spin up injection to the QSHI-SC-QSHI structure in the lower leg.}
	\label{fig:setup}
\end{figure}

\label{sec:introduction}
Many applications of quantum information require entanglement of quantum states \cite{Einstein1935,  Bennett2000, Nielsen2004}. While entanglement was achieved using photons \cite{Aspect1982, Zeilinger1999}, it is an ongoing challenge to create entangled electrons in solid state devices. S-wave superconductors (SC), which couple spin up and spin down electrons to Cooper pairs, could be used to provide spin entangled electrons by the inverse crossed Andreev reflection (CAR) \cite{Recher2001, Lesovik2001}, a process which splits a Cooper pair in the SC into two spatially separated, but spin-entangled, electrons in the normal region by the application of a bias voltage. Hence it is interesting to study the properties of the CAR and how its magnitude can be controlled. A straightforward way to observe CAR is by non-local conductance measurements \cite{Beckmann2004, Russo2005}. However, this method has the drawback that CAR is disguised by another non-local process called electron cotunneling \cite{Falci2001}, which does not involve Cooper pairs and is therefore a parasitic process. More involved experimental setups could recently detect Cooper-pair splitting circumventing the electron cotunneling processes, including additional quantum dots \cite{Hofstetter2010, Herrmann2010} or current noise \cite{Jian2010} measurements.
\\
In this paper we make use of the helicity conservation of the edge states in quantum spin Hall insulators (QSHI) \cite{Kane2005, Bernevig2006, Koenig2007Science} to achieve a spatial separation of the CAR from all other transport channels. Unlike previous works on QSHI-SC-QSHI interfaces \cite{Sato2010, Adroguer2010, Chen2011} we do not restrict ourselves to a phenomenological model of the edge states, but solve the full scattering problem within a generalized wave matching method \cite{Zhang2010, Reinthaler2012}. This method enables us to take into account a finite doping in the superconducting region, which is often the byproduct of the proximity effect. Exactly this mixing of the electron and hole scattering channels due to evanescent bulk modes in the SC results in the spatially localized CAR signal. Further we propose setups which allow for direct, all-electrical measurements of the CAR process, provided that only one spin direction contributes to the transport. The spin filter can be realized by contacting individual edges of the system separately (5-terminal setup, c.f. Fig.~\ref{fig:setupA}) or by using the non-equilibrium quantum spin Hall effect \cite{Bruene2011}, c.f. Fig.~\ref{fig:setupB}. Note, the measurement of CAR as well as the spin filter destroy the spin entanglement of the underlying Cooper pair.  However, spin-entangled electrons could be produced by inverse CAR processes in a setup where both edges are coupled to the same reservoirs.\\ 
Without excluding scattering channels, like it was done in p-n junctions \cite{Chen2011}, we can achieve CAR of up to about 50\% of the total non-local signal, which can be tuned by gating the system.

\label{sec:setup}
We consider junctions between two quantum spin Hall insulating leads and a superconductor (SC) of length $L$, as it is shown in Fig.~\ref{fig:setupA} together with the respective dispersions. The transport in the leads is characterized by helical edge states, c.f. the solid lines in the dispersion. When the wire has a finite width $W$, the linear dispersion of the edge states acquires a gap due to the overlap of the counter propagating edge states at opposite edges \cite{Zhou2008}. This so-called mini-gap is schematically shown in the electron dispersion for the left and right lead in Fig.~\ref{fig:setupA}. In the central region, a superconductor is placed on top of the quantum spin Hall insulator, which induces a superconducting gap and in general dopes the system, which is indicated by the energy shift $C_2$ in the central dispersion. As a consequence, the superconducting gap opens in the bulk of the system, whose energy dispersion is indicated by dashed lines in Fig.~\ref{fig:setupA}. In the leads, we will consider no intrinsic doping $C_{1,3}=0$ for simplicity. We will change the Fermi energy $\Ef$ in the whole structure which would imitate a back gate in the corresponding experimental setup \cite{Molenkamp2012}. Since the SC acts as a reservoir, its chemical potential is fixed and set to ground. \\
Let us assume a spin up electron is injected from the left lead, like shown in Fig.~\ref{fig:setup}. 
There are four spin conserving scattering mechanisms possible. The electron can be reflected (R) or tunnel through the sample (T) as an electron. To enter the superconducting condensate it needs a partner of opposite spin, which ejects a spin up hole either in the left or right lead. These processes are called local (AR) or crossed Andreev reflection (CAR), respectively. Due to the helicity of the edge states R and CAR are only allowed on the opposite edge, while AR and T take place on the edge of the incoming electron.	
Neglecting the doping due to the superconductor, i.e. $C_2=0$, Adroguer et al. \cite{Adroguer2010} found that in such superconducting tunneling junctions helicity conservation enforces perfect AR, when the width ($W$) and length ($L$) become large. Here we analyze effects of finite size and doping on the SC junction. The former effect introduces coupling of counter-propagating edge states and inter-lead tunneling due to narrow $W$ and short $L$ respectively. The latter effect results in the existence of gapped bulk states in the SC region. To tackle this problem, we use a generalized wave matching approach \cite{Zhang2010, Reinthaler2012}, which solves the full scattering problem within hard wall boundary conditions (HBC) \cite{noteHBC}. \\
Close to the $\Gamma$ point ($k=0$), HgTe quantum wells can be described by an effective $4\times4$ Dirac model \cite{Bernevig2006}. Using the basis $(\ket{E1+}, \ket{H1+}, \ket{E1-}, \ket{H1-})^T$, with (E1) and (H1) designating electron and heavy hole sub-bands and $\pm$ denoting the Kramers' partners, which we here call spin for simplicity, the Hamiltonian reads ($k_\pm = k_x \pm \imag k_y$, $k=\sqrt{k_x^2 +k_y^2}$, $k_\parallel = (k_x,k_y)^T$)
\begin{align}
	\label{eq:Hamiltonian}
	\mathcal H = \begin{pmatrix}
			h(k_\parallel) & 0 \\
			0 & h^*(-k_\parallel) \\
		\end{pmatrix}, \quad h(k_\parallel) = \ek \UnitM + d_{\alpha} (k_\parallel) \sigma_{\alpha},
\end{align}
with $d_\alpha (k_\parallel) = (Ak_x, -Ak_y, \Mk)^T$. $\sigma_\alpha$ are the Pauli matrices and $\UnitM$ is the unit matrix in the subband space.
Here $\ek=C-D k^2 $ is the energy dispersion, $\Mk = M - B k^2$ is the $k$ dependent bulk gap and $A$ gives the strength of the coupling between electron and heavy hole states. $C = C_2+E_f$ is an energy shift due to the doping by the superconductor ($C_2$) or due to a back gate ($E_f$). The parameters $A$, $B$, $D$ and $M$ depend on the quantum well width. We will treat the SC, which couples time reversed states, in the usual Bogoliubov-de Gennes (BdG) formalism. Since we neglected structure and bulk inversion symmetry breaking terms \cite{Rothe2010, noteBIA}, $\mathcal H$ consists of two decoupled blocks so that we can restrict our calculations to one spin direction. The influence on structure inversion asymmetry breaking terms to the proximity effect in helical edge states has been studied in Ref.~\cite{Virtanen2012}. We omit inter-edge Coulomb interaction which is smaller than all other energy scales \cite{Tanaka2009}. For the upper block, the BdG equation then reads \cite{Guigou2010}
\begin{align}
	\label{eq:BdG}
	  \mathcal H_{\text{BdG}}(-\imag \nabla_\parallel) 
	 \psi =\begin{pmatrix}
		h(-\imag \nabla_\parallel) & \Delta \UnitM\\
		\Delta^* \UnitM & - h(-\imag \nabla_\parallel)\\
	\end{pmatrix} \psi = \epsilon \psi,
\end{align} 
where we defined $\mathcal H_{\text{BdG}}$. Eq.~\eqref{eq:BdG} is correct in the absence of time reversal breaking potentials. $\psi= (u_{E1, +},u_{H1, +},v_{E1, -},v_{H1, -})^T $, where $u$ and $v$ are electron and hole like excitations. 
$\Delta$ is zero in the leads and constant in the central region \cite{note2} and $\epsilon$ is the excitation energy relative to the chemical potential in the SC. \\
We will consider transport in $x$-direction with HBC in $y$-direction. An analytical solution for HBC exists \cite{Zhou2008} for each region, but the scattering problem couples $y$-modes of different regions so that a simple wave matching 
is not possible. Here we use the orthogonal set $\phi_n(y) = \sqrt{2/W} \sin\left[n \pi y/W \right]$ ($n\in \mathbb N$) to Fourier expand the eigenstates in a certain part of the slab
\begin{align}
	\label{eq:FourierAnsatz}
	\psi_m (x,y)= \exp\left[\imag k_x^m x  \right] \sum_{n  = 1}^\infty \chi_{n}^m \phi_n(y),
\end{align}
where $m$ labels the different $k_x^m$ values. The $\chi_{n}^m$ are four component spinors in the basis used in Eq.~\eqref{eq:BdG}. The $k_x^m$ and the corresponding $\chi^m_n$ can be determined by inserting the ansatz \eqref{eq:FourierAnsatz} into the BdG Eq.~\eqref{eq:BdG} and rewriting the obtained system of equations as an eigenvalue equation for the $k_x^m$, see the appendix for a detailed discussion. 
While in the SC the electrons and holes are coupled, we can distinguish electrons from holes in the leads. 	 
\\
Having the solutions of the individual regions we can build up the scattering states for the QSHI-SC-QSHI junctions. In the following we will restrict ourselves to incoming spin up electrons from the left side, i.e. the upper edge state in Fig.~\ref{fig:setupA}. Since the interfaces break translational invariance and since the SC couples electrons and holes, the scattering states are combinations of solutions \eqref{eq:FourierAnsatz} with all allowed $k_x$, weighted by scattering amplitudes. The amplitudes are calculated by enforcing continuity of the wave function as well as the current at the interfaces. The transport coefficients R, T, AR and CAR can then be obtained from the absolute values squared of the amplitudes connecting in and out-going edge states, see e.g. Eqs~\eqref{eq:coeff1} and 
\eqref{eq:coeff2} in the appendix. These coefficients therefore are the probabilities that the incoming electron undergoes the corresponding scattering process. All the other coefficients couple to evanescent modes and do not contribute to transport in the leads. However, the transport by evanescent modes is crucial in the SC region.
\\
For the following calculations we have chosen the parameters from Ref.~\cite{Rothe2010}: $A = 0.3647\unit{nm\,eV}, \; B = -0.7061 \unit{nm^2 eV}, \; D = -0.5315 \unit{nm^2 eV}, \; M = -10.09\unit{meV}$ and $\Delta=0.5\unit{meV}$, which is a realistic value for the induced superconductivity by a s-wave SC, e.g. for bulk Nb $\Delta(\text{Nb}) \approx 1.45 \unit{meV}$ at zero temperature \cite{Chrestin1997}. We restrict our analysis to the $\epsilon=0$ regime.

\begin{figure}
	\subfloat[]{\includegraphics[width=0.75\linewidth]{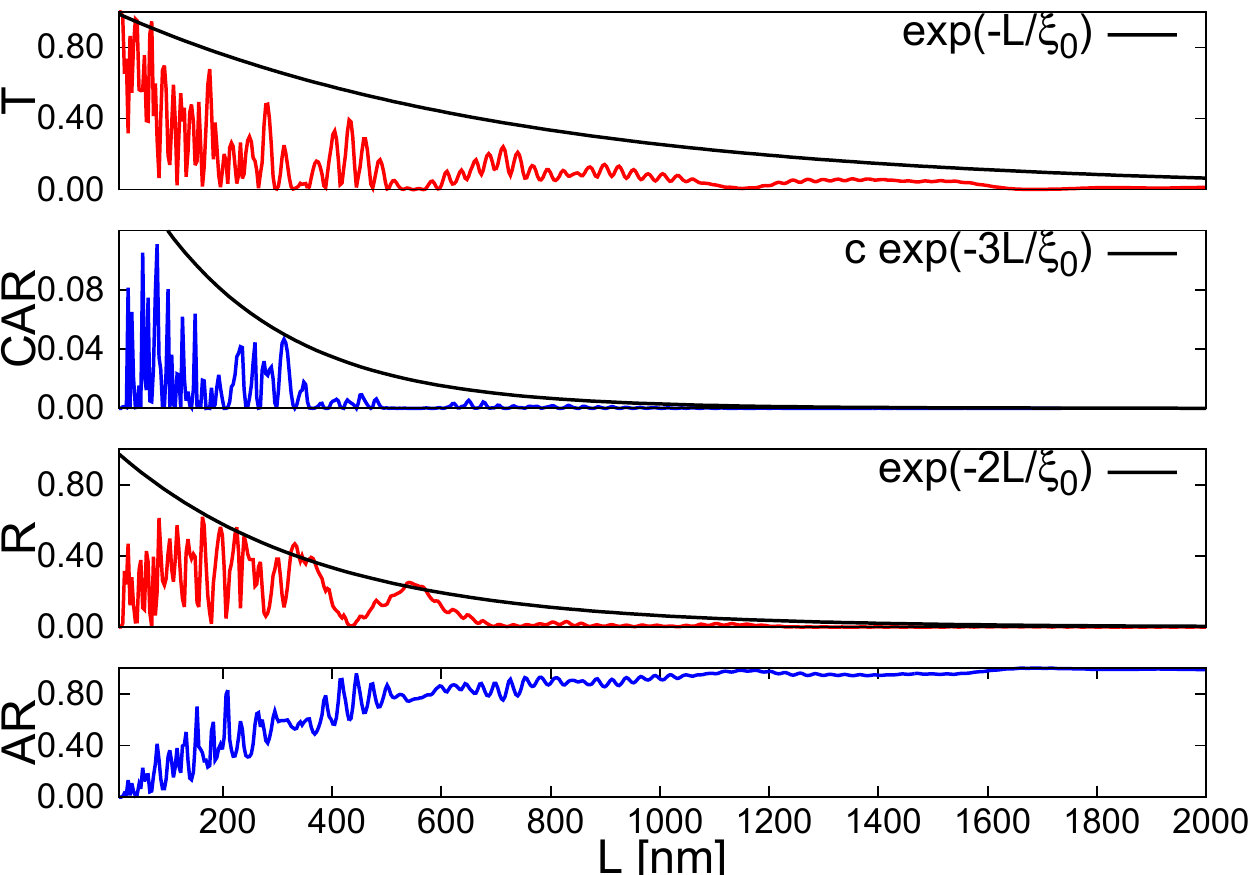}\label{fig:ARCTlength}}\\
	\subfloat[]{\includegraphics[width=0.75\linewidth]{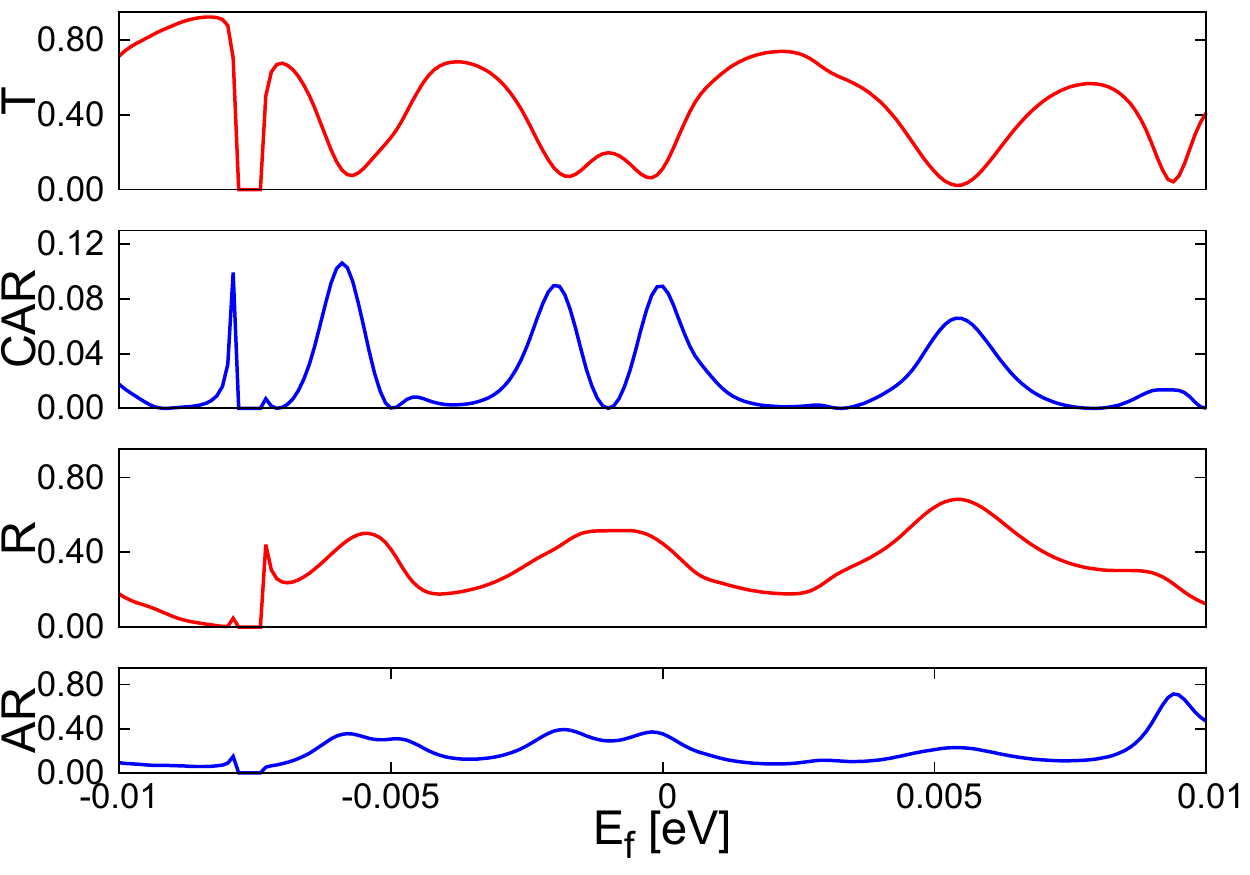}\label{fig:ARCTgate}}
	\caption{(a) The transport coefficients as a function of $L$ for $W=200\unit{nm}$. In addition to the resonant behavior we observe an exponential decay on the length scale of $\xi_0$. The constant $c=0.18$ was chosen to match the CAR signal. (b) The transport coefficients as a function of the global doping $\Ef$ at fixed geometry $W=200\unit{nm}$ and $L=100\unit{nm}$.}
	\label{fig:fig3}
\end{figure}

\label{sec:results}
\begin{figure}
	\subfloat[]{\includegraphics[width=0.75\linewidth]{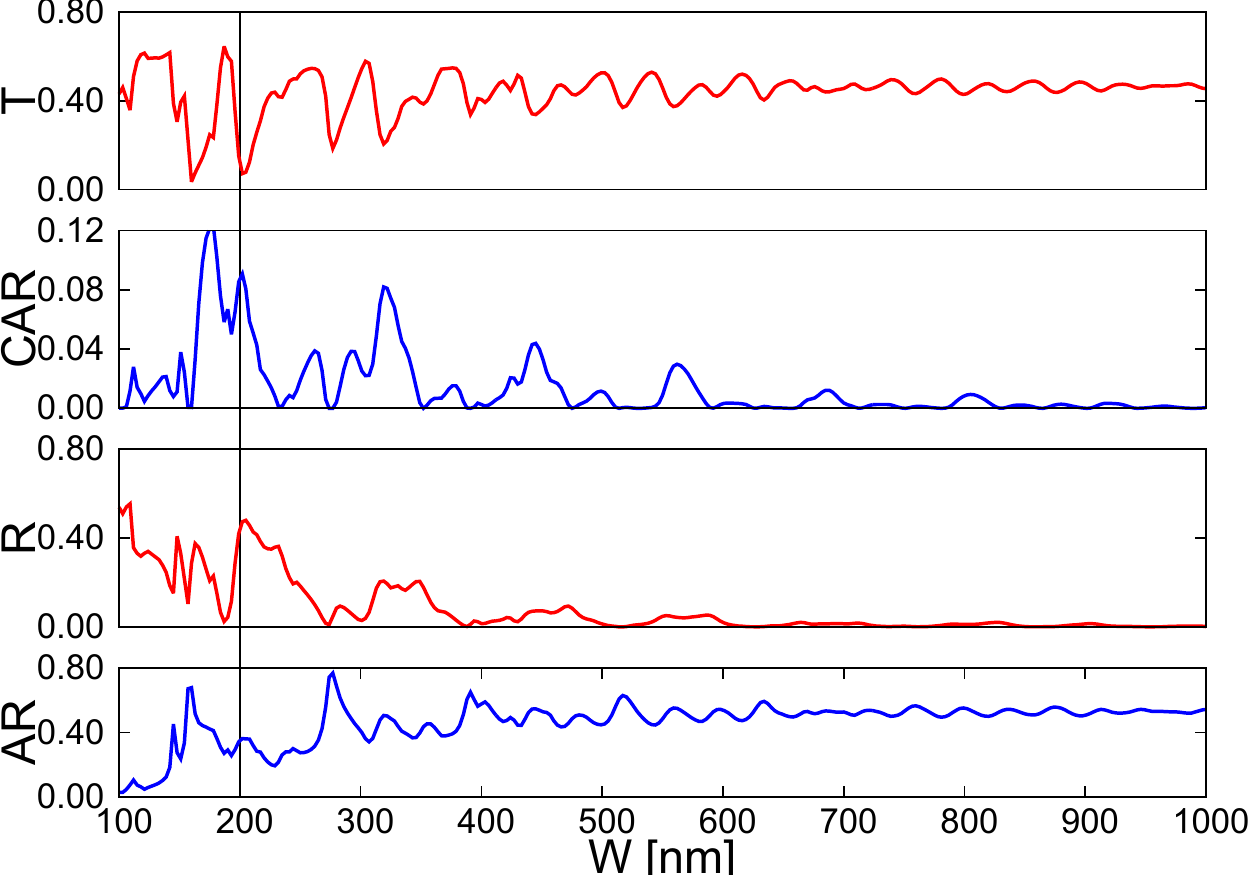}\label{fig:ARCTwidth}}\\
	\vspace{-1cm}
	\subfloat[]{\includegraphics[width=0.85\linewidth]{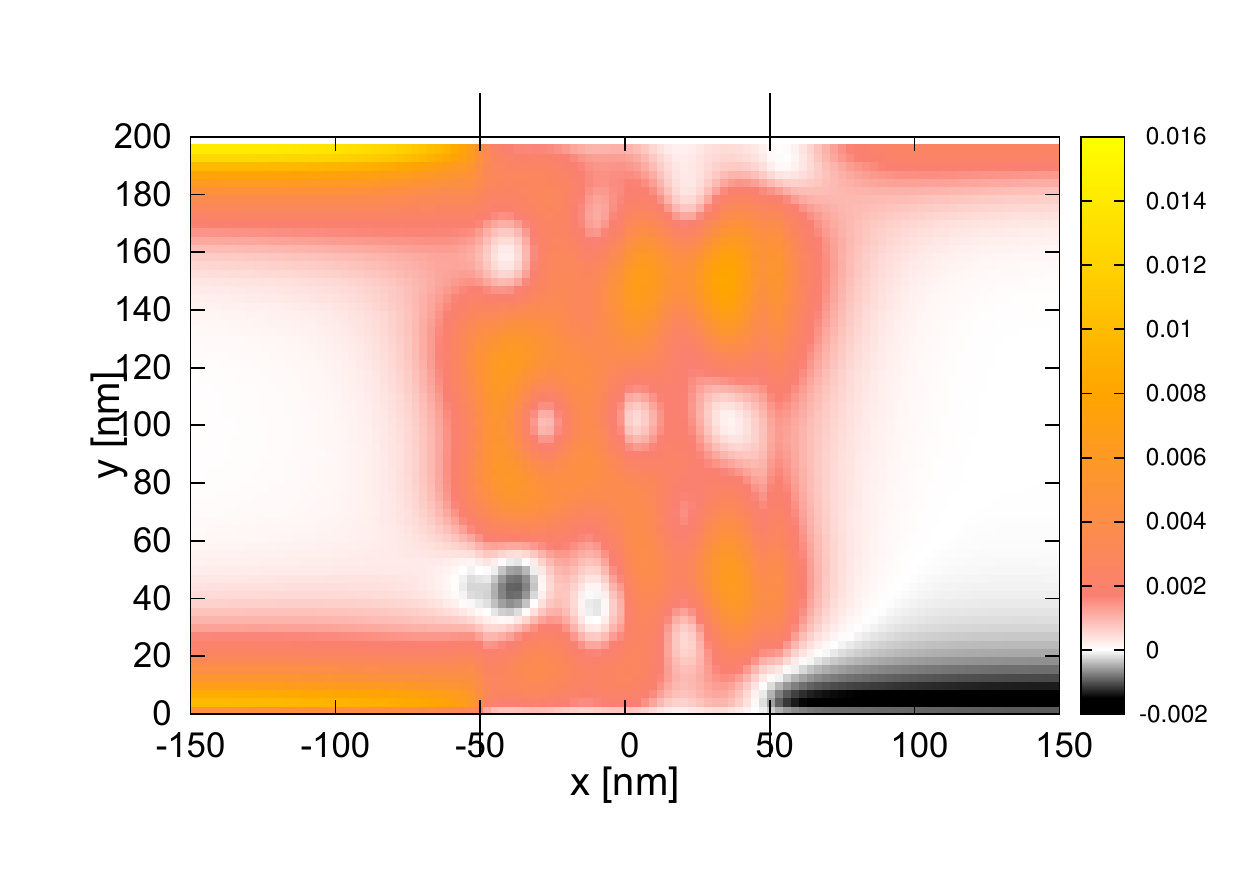}\label{fig:density}}
	\caption{(a) The transport coefficients as a function of the width for length $L=100\unit{nm}$, $C_1 = C_3 = 0$ and $C_2 = -50 \unit{meV}$. 
We see that the finite size effects increase for smaller width, while at large width CAR and R are suppressed. (b) The charge density at width $200\unit{nm}$ (cf. black line in (a)), i.e. at large CAR. 
Gray (warm) colors and positiv numbers correspond to electronlike states, black and negative numbers to an excess of holes. 
We see that the incoming electron edge state (left lead upper edge) is scattered through the SC bulk modes into a spatially separated cotunneled electron and the CAR hole edge state on the right side.}
	\label{fig:fig2}
\end{figure}
In Fig.~\ref{fig:ARCTwidth} we show the transport coefficients as a function of $W$ for a constant $L=100\unit{nm}$. We find that R and CAR approach zero for $W \gtrsim 600\unit{nm}$, which is in agreement with Ref.~\cite{Adroguer2010}, although $C_2 = -50\unit{meV} $ dopes the system far outside the bulk gap. However, we do not find quantized AR, since $L$ is relatively small and electrons can tunnel through the junction. Increasing $L$ restores perfect AR. In contrast to Ref.~\cite{Adroguer2010} we are interested in finite size effects which dominate at $W <  600\unit{nm}$. In particular, we observe large fluctuations in the signal as well as finite R and CAR. The latter reaches its maxima at points, where T is relatively small. Therefore it is possible that CAR and T have the same strength of about 10\% of the total signal in this configuration. At these points the charge current in the right lead becomes minimal and spin currents dominate the transport \cite{Das2008}.
Fig.~\ref{fig:density} shows the charge density for $W=200\unit{nm}$. Gray (warm) colors and positiv numbers correspond to electronlike states, black and negative numbers to an excess of holes. The two interfaces at $-L/2$ and $L/2$ are indicated by vertical black lines. Inside the leads we find finite density only along the edges. On the left side we have an incoming electron, which overlaps with a locally reflected hole, and a reflected electron on the other edge. The incoming electron is either directly reflected as a hole (AR) or enters the SC. No direct electron reflection happens at the first interface \cite{Reinthaler2012}
. Inside the SC the electron is reflected several times, which explains the resonant behavior at small width. Since the electron solution is evanescent for $0=\epsilon<\Delta$, the resonant behavior due to multiple reflections decays with increasing width. A rough estimate for the scale on which the resonant behavior can be observed can be obtained by the SC coherence length $\xi_0 = \hbar v_F/\Delta$. For small energies, quadratic terms in \eqref{eq:Hamiltonian} play a minor role \cite{Schmidt2009} and the approximation $\xi_0 \approx A/\Delta = 729\unit{nm}$ holds and all resonant behavior should take place at $W\lesssim\xi_0$. In Fig.~\ref{fig:ARCTwidth}, R and CAR are suppressed earlier, which is due to the fact that the particle does not take the direct way, but traverses the sample several times. In the right lead the solutions of T on the upper edge and of CAR on the lower edge are spatially separated due to helicity conservation. The separation works, as long as the edge states, which decay like $\exp[-\lambda y]$ away from the edges, do not overlap. This is fulfilled for our setup since from the parameters of our modelling we find $1/\lambda \approx 20\unit{nm} \ll W$. This separation allows for a direct observation and electrical measurement of the CAR. Before we describe the proposed detection schemes, let us analyze the occurrence of finite CAR in greater detail.\\
In Fig.~\ref{fig:ARCTlength} we show the transport coefficients as a function of the SC's length $L$ at fixed $W=200\unit{nm}$.	
Here we observe that for long $L$ the probability of AR goes to one. The other transport coefficients decay due to the evanescent nature of the particles. The decay lengths can be fitted by the coherence length $\xi_0$ (T), $\xi_0/2$ (R) and $\xi_0/3$ (CAR). This indicates that the particle has to traverse the sample several times in order to contribute to R or CAR. 
The need of additional scattering events at the interfaces to couple states at opposite edges shows that the edge states are not coupled by direct overlap, which is suppressed at large width. Instead the existence of evanescent bulk modes, which span over the whole sample, allows for finite R and CAR. At $C_2=0$, the absence of weakly decaying bulk modes prohibits transitions between the different edges which can then only be introduced by the direct overlap of edge states. However, this overlap would destroy the spatial separation of CAR and T in the right lead.\\
On top of the evanescent decay we find resonant oscillations of the signal. The latter can be understood in the context of Fabry-P\'erot oscillations. As long as $L \ll \xi_0$ is fulfilled, the signal oscillates on length scales of $ L= \pi/\Re(k^m_x)$. However, since many different bulk modes contribute, the periodicity can not be quantified in general. Nevertheless, instead of changing the geometry, which is experimentally difficult, the Fabry-P\'erot condition can be fulfilled for a fixed $L$ by varying $\Re(k^m_x)$. Since $k^m_x =k^m_x(\Ef)$ we can use gates to realize maximal CAR for further investigation. Since the superconductor screens the system, gating can be realized most efficiently by back gates. In Fig.~\ref{fig:ARCTgate} we present the behavior when the back gate changes the Fermi energy in the whole system. For the energy range shown, the leads are not doped outside of the bulk gap, which ensures edge state transport and spatial separation of T and CAR. The signal shows the same kind of oscillations like when we changed $L$. Additionally we observe that the signal breaks down around $\Ef \approx -7.5\unit{meV}$. The mini-gap opens around this energy in the leads, because the overlap of the edge states changes as a function of the energy \cite{Zhou2008}. Close to the mini-gap we find a peak in CAR which is generated by the direct overlap of the edge states and hence does not allow for spatial separation of the transport signal, as mentioned before.\\
For the calculations we restricted ourselves to one spin direction, which is crucial for the spatial separation. Here we propose two different ways to detect the separation between T and CAR: It is possible to contact the edges independently in a 5-terminal setup \cite{Roth2009}, like it is shown in Fig.~\ref{fig:setupA}. Grounding all but contact 1, where a voltage $V_1$, is applied, drives only spin up electrons towards the junction. Non-local conductance measurements, i.e. $\partial I_2/\partial V_1$ and $\partial I_3/\partial V_1$, then allow for a direct electrical measurement of T and CAR, respectively. In Fig.~\ref{fig:setupB} we propose to use the non-equilibrium quantum spin Hall effect in an H-bar structure for spin selection. Applying a current $I$ in the upper leg
drives electrons from right to left in the upper leg. Due to helicity only spin up electrons propagate along the lower edge of the upper leg. These spin up electrons are transmitted through the bridge to the lower leg, where they are injected to the junction and scattered in the R, AR, T or CAR edge states.

\label{sec:conclusions}
In conclusion, we propose that the CAR can be measured all-electrically in a QSHI - SC - QSHI junction. We presented transport calculations showing that finite size effects combined with doping the SC by the proximity effect enable a significant CAR, which is spatially separated from all other transport channels. We provided a handle to tune the magnitude of the signal by use of a back gate for given sample geometries. 
At points where T and CAR have the same magnitudes the non-local charge currents cancel and we find a pure spin current in the right lead without relying on ferromagnetic elements. 
We also argued, that for actual measurements spin selection can be achieved by contacting individual edge states or by use of the quantum spin Hall effect. Our results should be equally relevant to other 2D topological insulators, like graphene \cite{Kane2005, Weeks2011} and thin films of Bi$_2$Se$_3$ \cite{Liu2010a}.

\label{sec:acknowledgements}
We thank L.W. Molenkamp and J. Cayssol for helpful discussions. We acknowledge financial 	ort by the DFG grants HA 5893/3-1, RE 2978/1-1 and from the EU FP7 project SE2ND [271554].


\begin{thebibliography}{40}
\bibitem[{not({\natexlab{a}})}]{noteHBC}
\bibinfo{note}{The exact form of the boundary conditions is not important as
  long as there is a domain wall between topologically trivial and non-trivial
  phases.}
\bibitem[{not({\natexlab{b}})}]{noteBIA}
\bibinfo{note}{The found separation effect is robust for a bulk inversion
  asymmetry $\lesssim 1 \; \mathrm{meV}$, which is the experimentally relevant
  coupling strength \cite{Molenkamp2012}.}
\bibitem[{not({\natexlab{c}})}]{note2}
\bibinfo{note}{The form of the $\Delta$-matrix in $(E1,H1)$-space (proportional
  to the unit-matrix) in Eq. \eqref{eq:BdG} is a simplification. However, in
  the regime considered, the bulk eigenstates of Eq.~\eqref{eq:Hamiltonian} are
  a mixture of $E1$ and $H1$ bands so that we only find quantitative changes in
  the different scattering channels for a more general $\Delta$-matrix.}
\end{thebibliography}

\appendix*
\section{Details on the method}
\label{app:A}
Here we follow the detailed discussion of the method (see also Ref.~\cite{Reinthaler2012}). 
Putting the Fourier ansatz into the BdG equation (see Eq.~\eqref{eq:FourierAnsatz} and Eq.~\eqref{eq:BdG} in the main text respectively) and multiplying from the left by $\int_0^W \rd y \phi^\dagger_{n_2}(y)$ we arrive at 
\begin{align}
	\label{eq:EigenvalueProblemNoMat}
  \mathcal{H}^{\text{const}} \chi^m_{n_2} &+ \mathcal{H}^{k_x} k^m_x \chi^m_{n_2}  + \mathcal{H}^{k_x^2} k^m_x \chi^{\prime m}_{n_2} \nonumber \\
  &+ \sum_{n_1}\int_0^W \rd y\phi_{n_2}^\dagger \mathcal{H}^{k_y} \phi_{n_1} \chi^m_{n_1} = 0 ,
\end{align}
where all terms of the Hamiltonian proportional to $k_y = -\imag \partial_y$
 are collected in $\mathcal{H}^{k_y}$, all constant terms in $ \mathcal{H}^{\text{const}}$ and all with $k_x$ ($k_x^2$) in $ \mathcal{H}^{k_x}$ ($\mathcal{H}^{k_x^2}$). Additionally we introduced $\chi_{n}^{\prime m} = k_x^m \chi_n^m$. Defining the vectors $\chi^m = (\chi^m_{n=1}, \chi^m_{n=2}, \ldots)^T$ and $\chi^{\prime m} = (\chi^{\prime m}_{n=1}, \chi^{\prime m}_{n=2}, \ldots)^T$, which are built from the $4$ component spinors $\chi_n^m$ and $\chi_n^{\prime m}$, Eq.~\eqref{eq:EigenvalueProblemNoMat} can be written as a matrix equation
\begin{align}
	\label{eq:EigenvalueProblem}
	\begin{pmatrix}
		\UnitM & 0\\ 0 & \left(H^{k_x^2}\right)^{-1}\\
	\end{pmatrix}
	\begin{pmatrix}	
		0 & \UnitM \\ H^{\text{const}} + H^{k_y} & H^{k_x} \\
	\end{pmatrix} 
	\begin{pmatrix} \chi^m \\ \chi^{\prime m} \end{pmatrix}
		=
	k_x^m 	
	\begin{pmatrix} \chi^m \\ \chi^{\prime m} \end{pmatrix}.
\end{align}
So doubling the dimension of the system of equations by introducing $\chi^{\prime m}$ allows us to reduce the problem of finding $k_x^m$ to a linear eigenvalue equation \cite{Chang1982, Liu2011}. The sub-matrices in Eq.~\eqref{eq:EigenvalueProblem} are 
\begin{align}
	H^{k_x^2}_{n_1n_2} &= \delta_{n_1n_2} 
	\mathrm{diag} \left[\tilde{D}_+,\tilde{D}_- , -\tilde{D}_+ , -\tilde{D}_-\right],\\
   H^{k_x}_{n_1n_2} &=\delta_{n_1n_2}
  \begin{pmatrix}
		0 & A & 0 & 0\\
		A & 0 & 0 & 0\\
		0 & 0 & 0 &-A\\
		0 & 0 &-A & 0\\
	\end{pmatrix}, \\
	H^{\text{const}}_{n_1n_2} &=\delta_{n_1n_2} \begin{pmatrix}
													\tilde{C}_+ - \epsilon & 0 & \Delta & 0\\
													0 & \tilde{C}_- - \epsilon & 0 & \Delta\\
													\Delta^*& 0 & -\tilde{C}_+ -\epsilon & 0 \\
													0 & \Delta^*& 0 &-\tilde{C}_- -\epsilon\\
												\end{pmatrix},\\	
	H^{k_y}_{n_1n_2} &=  
	\begin{pmatrix}
		- \kappa_{n1,n2}\tilde{D}_+  & \imag A \eta_{n_1n_2}  & 0 & 0\\
		-\imag A \eta_{n_1n_2} & -\kappa_{n1,n2}\tilde{D}_- & 0 & 0\\
		0 & 0 & \kappa_{n1,n2}\tilde{D}_+& -\imag A \eta_{n_1n_2} \\
		0 & 0 & \imag A \eta_{n_1n_2} &\kappa_{n1,n2}\tilde{D}_-\\
    \end{pmatrix}.
\end{align}
Further we used here $\tilde{D}_\pm = (D \pm B)$, $\kappa_{n1,n2} = \left(\frac{n \pi }{W}\right)^2\delta_{n_1n_2}$ and $\tilde{C}_\pm = C\pm M$. The only term coupling different modes is $\eta_{n_1n_2} = \bra{\phi_{n_1}} k_y \ket{\phi_{n_2}}$. For actual computation the Fourier series has to be truncated at a sufficiently high mode $N$.\\
The solutions $k_x^m$ can be characterized as propagating ($k_x^m \in \mathbb{R}$) or evanescent ($\Im(k_x) \neq 0$).
For real $k_x^m$ the propagation direction can be determined by the sign of 
\begin{align}
	\label{eq:velocity}
v^m = \int_0^W \rd y \psi_m^\dagger (x,y)\left[\partial_{k_x} \mathcal H_{\text{BdG}} (k)\right]_{k_x \rightarrow k_x^m}  \psi_m (x,y).
\end{align} 
Analogously evanescent states with $\Im(k_x)>0$ ($\Im(k_x)<0$) are decaying to the right (left). While in the superconductor the electrons and holes are coupled we can distinguish electrons from holes in the leads. 
In the following we will denote right (left) propagating/decaying solutions by the index $m^i_R$ ($m^i_L$), where $i=e,h$ for electrons and holes in the leads. 
Equipped with this we can write the scattering states in each region. To do so we label the $k_x$ eigenvalues as:
$k_{x,1}^{m, i}$ for the part 1 stretching from $x = - \infty$ to $x = - L/2$, $k_{x,2}^{m}$ for $-L/2<x<L/2$ and $k_{x,3}^{m, i}$ for $x \ge L/2$, where $i=e,h$ gives the particle character in the leads. Our calculations will be restricted to zero temperature, where only incoming electrons (not holes) need to be considered. In the case where an electron is incoming in mode $m$ the states take the form:
\begin{align}
	\Psi_1 (x,y) =& \psi_{m,1}(x,y) + \sum_{m^e_L} r_{m^e_L, m} \psi_{m^e_L,1}(x,y) \nonumber \\ &+\sum_{m^h_L} r_{m^h_L, m} \psi_{m^h_L,1}(x,y),\\
	\Psi_2(x,y) =& \sum_{m_R}c_{m_R,m} \psi_{m_R,2} (x,y) +\sum_{m_L} d_{m_L, m} \psi_{m_L,2}(x,y), \\
	\Psi_3(x,y) =& \sum_{m^e_R} t_{m^e_R,m} \psi_{m^e_R,3}(x,y)+ \sum_{m^h_R} t_{m^h_R,m} \psi_{m^h_R,3}(x,y).
\end{align}
Here $\vert r_{m^e_L, m}\vert^2 $ ($\vert r_{m^h_L, m}\vert^2 $) and $\vert t_{m^e_R,m}\vert^2$ ($\vert t_{m^h_R,m}\vert^2$) are the probabilities that the incoming mode $m$ is reflected into the electron in mode $m^e_L$ (hole in mode $m^h_R$) of the left lead or transmitted into mode $m^e_R$ ($m^h_R$) of the right lead, respectively. Enforcing continuity of the wave functions and of the currents $\left[\nabla_{\vec{k}} H_{\text{BdG}}(k) \right]\Psi(x,y)$ allows to calculate all scattering coefficients. In the setup of Fig.~\ref{fig:setup} in the main text only edge state modes are propagating. The evanescent waves in the leads do not enter the scattering matrix. Let $m^i_D$ from now on be an edge state mode, which propagates in $D$ direction. Then we have:
\begin{align}
	\label{eq:coeff1}
	T =& \vert t_{m^e_R, m} \vert^2 \frac{v^{m^e_R}}{v^m}, \quad R = \vert r_{m^e_L, m} \vert^2 \frac{v^{m^e_L}}{v^m} \\
	\label{eq:coeff2}
	CAR =&\vert t_{m^h_R, m} \vert^2 \frac{v^{m^h_R}}{v^m}, \quad AR = \vert r_{m^h_L, m} \vert^2 \frac{v^{m^h_L}}{v^m}.
\end{align}
Using these transport coefficients, we can express the local and non-local conductance $G_{ij}(\epsilon) = \rd I_i/\rd V_j$, i.e. the conductance from contact $j$ to $i$, where $i,j = 1,\ldots,4$ corresponds to the contacts in Fig.~\ref{fig:setupA} in the main text. $\epsilon = e V_1$ is the voltage applied at contact 1. Assuming spatially well separated transport signals, the conductance at zero temperature can be approximated by \cite{Blonder1982}
\begin{align}
	G_{11} (\epsilon) &= \frac{e^2}{h}(1 + AR (\epsilon))\\
	G_{21} (\epsilon) &= \frac{e^2}{h}T (\epsilon) \\
	G_{31} (\epsilon) &= - \frac{e^2}{h}CAR (\epsilon) .
\end{align}
Moreover, the knowledge of all scattering amplitudes determines the full state $\Psi(x,y)$ up to normalization. $\Psi(x,y)$ is the response to an incoming mode $\psi_{m,1}(x,y)$ from the left lead. Therefore it allows us to calculate the non-equilibrium charge density $n (x,y) = \Psi^\dagger(x,y)\Lambda\Psi(x,y), \; \Lambda = \mathrm{diag}[1,1,-1,-1]$. For plotting the charge density we use the following normalization:
 \begin{align}
 	\int_0^W \rd y \Psi^\dagger (x = -\frac{L}{2}, y)\Psi (x = -\frac{L}{2}, y) = 1.
 \end{align}

\end{document}